\documentclass[preprint,12pt]{elsarticle}

\usepackage{epsfig}
\usepackage{amssymb}
\usepackage{amsthm}
\usepackage{amsmath}

\usepackage{lineno}

\journal{Nuclear Physics B}

\begin{document}

\begin{frontmatter}

\title{Double quarkonium production at high Feynman-$x$}
\author{Sergey Koshkarev, Stefan Groote}
\address{Institute of Physics, University of Tartu, Tartu 51010, Estonia}

\begin{abstract}
In this paper we give estimates for the proton--proton cross sections into
pairs of quarkonium states $J/\psi$, $\psi(2S)$, $\Upsilon(1S)$ and
$\Upsilon(2S)$ at the scheduled AFTER@LHC energy of $115$~GeV. The estimates
are based on the intrinsic heavy quark mechanism which is observable for high
values of $x_F$, a range outside the dominance of single parton and double
parton scattering.
\end{abstract}

\end{frontmatter}


\section{Introduction}
\label{intro}

In the era of high luminosity and high energy accelerators the associated
heavy quarkonium production plays a special role as a testing ground to study
multiple parton scattering in a single hadron collision. Significant progress
on the Double Parton Scattering (DPS) has been provided by the Tevatron and
the LHC in measuring the productions of $J/\psi + W$~\cite{ATLAS},
$J/\psi + Z$~\cite{ATLAS2}, $J/\psi$ + charm~\cite{LHCb} and
$J/\psi + J/\psi$~\cite{LHCb_psi,DZero,CMS}. Therefore and for many other
reasons, heavy quarkonium production is always a hot topic in high energy
physics, as this kind of physics is an ideal probe for testing quantum
chromodynamics. 

Current colliders provide access only to the physics at low values of the
Feynman parameter $x_F$. However, significant interest is given also for
physics at high $x_F$~\cite{Brodsky2013,Lansberg2012,Lansberg2014,%
Rakotozafindrabe,Brodsky2015}. This region will be accessible at a future
fixed-target experiment at the LHC (\hbox{\rm AFTER@LHC}). In a recent paper,
Jean-Philippe Lansberg and Hua-Sheng Shao discussed contributions of the DPS
to the double-quarkonium production in the kinematic region of the
\hbox{\rm AFTER@LHC}~\cite{Shao}. However, as we learned from the low
statistics NA3 experiment measurements of the double $J/\psi$
production~\cite{NA31982,NA31985} and the observation of the double charmed
baryons by the SELEX collaboration~\cite{SELEX2002,Mattson,SELEX2005}, the
double intrinsic heavy quark mechanism can be the leading production
mechanism~\cite{Vogt1995,Koshkarev}.

The existence of a non-perturbative intrinsic heavy quark component in the
nucleon is a rigorous prediction of QCD. Intrinsic charm and bottom quarks are
contained in the wavefunction of a light hadron  --  from diagrams where the
heavy quarks are multiply attached via gluons to the valence quarks. In
detail, the intrinsic heavy quark components are contributed by the twist-six
contribution of the operator product expansion proportional to
$1/m_Q^2$~\cite{Brodsky1984,Franz}. In this case, the frame-independent
light-front wavefunction of the light hadron has maximum probability if the
Fock state is minimally off-shell. This means that all the constituents are at
rest in the hadron rest frame and thus have the same rapidity $y$ if the
hadron is boosted. Equal rapidity occurs if the light-front momentum fractions
$x=k^+/P^+$ of the Fock state constituents are proportional to their
transverse masses, $x_i \propto m_{T, i} = ( m^2_i + k^2_{T,i} )^{1/2}$, i.e.\
if the heavy constituents have the largest momentum fractions. This features
the BHPS model given by Brodsky, Hoyer, Peterson and Sakai for the
distribution of intrinsic heavy quarks~\cite{Brodsky1980,Brodsky1981}.

In the BHPS model the wavefunction of a hadron in QCD can be represented as a
superposition of Fock state fluctuations, e.g.\ $| h \rangle \sim | h_l
\rangle + | h_l g \rangle + | h_l Q \bar{Q} \rangle \ldots$, where $ h_l$ is
the light quark content, and $Q=c,b$. If the projectile interacts with the
target, the coherence of the Fock components is broken and the fluctuation can
hadronize. The intrinsic heavy quark Fock components are generated by virtual
interactions such as $gg \to Q \bar{Q}$ where the gluons couple to two or more
valence quarks of the projectile. The probability to produce such $ Q \bar{Q}$
fluctuations scales as $\alpha_s^2 (m_Q^2)/m_Q^2$ relative to the
leading-twist production.

Following Refs.~\cite{Vogt1995,Brodsky1980,Brodsky1981}, the general formula
for the probability distribution of an $n$-particle intrinsic heavy quark
Fock state as a function of the momentum fractions $x_i$ and the transfer
momenta $\vec{k}_{T,i}$ can be written as
\begin{equation}
\begin{aligned}
\frac{dP_{iQ}}{\prod_{i=1}^n dx_i d^2 k_{T,i}} \propto
  \alpha_s^4 (M_{Q \bar Q}) \frac{\delta\big(\sum_{i=1}^n \vec{k}_{T,i}
  \big) \delta \big( 1 - \sum_{i=1}^n x_i \big)}{\big( m_h^2 -
  \sum_{i=1}^n m_{T,i}^2 / x_i \big)^2},
\end{aligned}
\end{equation}
where $m_h$ is the mass of the initial hadron. The probability distribution
for the production of two heavy quark pairs is given by
\begin{equation}
\begin{aligned}
\frac{dP_{iQ_1Q_2}}{\prod_{i=1}^n dx_i d^2 k_{T,i}} \propto
  \alpha_s^4 (M_{Q_1 \bar{Q}_1}) \alpha_s^4 (M_{Q_2 \bar{Q}_2})
  \frac{\delta \big(\sum_{i=1}^n \vec{k}_{T,i} \big)\delta
  \big( 1 - \sum_{i=1}^n x_i \big)}{\big( m_h^2 -  \sum_{i=1}^n
  m_{T,i}^2 / x_i \big)^2}.
\end{aligned}
\end{equation}
If one is interested in the calculation of the $x$ distribution, one can
simplify the formula by replacing $m_{T,i}$ by the effective mass
$\hat{m}_i = ( m_i^2 + \langle k^2_{T,i} \rangle )^{1/2}$ and neglecting the
masses of the light quarks,
\begin{equation}
\begin{aligned}
 \frac{dP_{iQ_1Q_2}}{\prod_{i=1}^n dx_i} \propto
  \alpha_s^4 (M_{Q_1 \bar{Q}_1}) \alpha_s^4 (M_{Q_2 \bar{Q}_2})
  \frac{\delta \big( 1 - \sum_{i=1}^n x_i \big)}{\big(\sum_{i=1}^n
  \hat{m}_{T,i}^2 / x_i \big)^2}.
\end{aligned}
\end{equation}
The $x_F$ distribution for the double quarkonium production $X_1+X_2$
(with $X_i=J/\psi,\psi(2S),\Upsilon(1S),\Upsilon(2S),\ldots$) is then
given by~\cite{Vogt1995}
\begin{eqnarray}
\frac{dP_{iQ_1Q_2}}{dx_{X_1X_2}}&=&\int\prod_{i=1}^n dx_idx_{X_1}dx_{X_2}
  \frac{dP_{iQ_1Q_2}}{\prod_{i=1}^n dx_i}\delta(x_{X_1X_2}-x_{X_1}-x_{X_2})
  \nonumber\\&&\times\delta(x_{X_1}-x_{Q_1}-x_{\bar Q_1})
  \delta(x_{X_2}-x_{Q_2}-x_{\bar Q_2}).\qquad
\end{eqnarray}
The BHPS model assumes that the vertex function in the intrinsic heavy quark
wavefunction is varying relatively slowly. The particle distributions are then
controlled by the light-cone energy denominator and the phase space. The Fock
states can be materialized by a soft collision in the target which brings the
state on shell. The distribution of produced open and hidden charm states will
reflect the underlying shape of the Fock state wavefunction.

In this paper we investigate the double intrinsic heavy quark mechanism for
the double-quarkonium production in the high Feynman-$x$ region at the
\hbox{\rm AFTER@LHC} experiment. In this particular case the production of the
double quarkonium plays a special role as it provides the direct access to
extract the double heavy quark probabilities $P_{icc}$, $P_{icb}$ and
$P_{ibb}$. To the best of our knowledge the $x_F$ distribution for
double-quarkonium production in proton beam events has not yet been measured
(cf.\ also a comment at the end of the third paragraph in the Introduction
of Ref.~\cite{Vogt1995}). Therefore, our estimates cannot be compared to
existing data but wait for future confirmation by experiments like
\hbox{\rm AFTER@LHC}, for which we give numerical values. As an innovative
element, for our analysis we use the color evaporation model, applied also to
excited $2S$ states. Finally, in the conclusions we discuss why existing LHC
measurements cannot be interpreted as non-evidence of the intrinsic heavy
quark mechanism.

\section{Double-quarkonium production cross section\label{prod}}

The production cross section of the quarkonium can be obtained as an
application of the quark--hadron duality principle known as color evaporation
model (CEM)~\cite{Amundson}. In this model the cross section of quarkonium are
obtained by calculating the production of a $Q \bar Q$ in the small invariant
mass interval between $2m_Q$ and the threshold to produce open heavy-quark
hadrons, $2m_H$. The $Q \bar Q$ pair has $3 \times \bar{3} = (1 + 8)$ color
components, consisting of a color-singlet and a color-octet. Therefore, the
probability that a color-singlet is formed and produces a quarkonium state is
$1/(1 + 8)$, and the model predicts
\begin{eqnarray}
\sigma(Q\bar Q) = \frac{1}{9} \int_{2m_Q}^{2m_H} dM_{Q \bar Q}
  \frac{d\sigma_{Q \bar Q}}{dM_{Q \bar Q}} = \frac{1}{9} \int_{4m_Q^2}^{4m_H^2}
  dM^2_{Q \bar Q} \frac{d\sigma_{Q \bar Q}}{dM^2_{Q \bar Q}}\, ,
\end{eqnarray}
where $\sigma_{Q \bar Q}$ is the production cross section of the heavy quark
pairs and $\sigma(Q\bar Q)$ is a sum of production cross sections of all
quarkonium states in the duality interval. For example, in case of charmonium
states one has $\sigma(Q\bar Q)=\sigma(J/\psi)+\sigma(\psi(2S))+\ldots$\,.
According to a simple statistical counting, the fraction of the total
color-singlet cross section into a quarkonium state is given by
\begin{equation}\label{XQQ}
\sigma(X) = \rho_X \cdot \sigma(Q\bar Q)
\end{equation}
($X=J/\psi,\psi(2S),\ldots$) with
\begin{equation}
\rho_X = \frac{2 J_X + 1}{\sum_i (2J_i + 1)}\, ,
\end{equation}
where $J_X$ is the spin of the quarkonium state $X$ and the sum runs over all
quarkonium states. In case of the $J/\psi$ meson the calculation gives
\begin{equation}\label{rhoJpsi}
\rho_{J/\psi} \simeq 0.2.
\end{equation}

This statistical counting rule works well for $J/\psi$ but not so well for
other charmonium states, even not for $\psi(2S)$. Instead, in this paper we
use the fact that a quarkonium production matrix element is proportional to
the absolute square of the radial wave function at the origin~\cite{Humpert},
so that
\begin{equation}
\sigma(J/\psi) : \sigma(\psi(2S)) \approx
  |R_{J/\psi}(0)|^2 : |R_{\psi(2S)}(0)|^2.
\end{equation}
The absolute square of the radial wave function $R_X(0)$ of the quarkonium
state $X=J/\psi,\psi(2S),\ldots$ at the origin is determined by the leptonic
decay rate~\cite{Eichten}
\begin{equation}
\Gamma(X\to e^+ e^-) = \frac{4N_c\alpha^2_{\rm em}e^2_Q}{3}
\frac{|R_X(0)|^2}{M_X^2}\bigg( 1 - \frac{16\alpha_s}{3\pi} \bigg),
\end{equation}
where $N_c = 3$ is the  number of quark colors, $e_Q$ is the electric
charge of the heavy quark, and $M_X$ is the mass of the quarkonium state $X$.
Splitting $\sigma(Q\bar Q)$ up into the different quarkonium states one can
obtain the corresponding production cross sections.

According to the intrinsic heavy quark mechanism the production cross section
$\sigma(Q\bar Q)$ of a $Q \bar Q$ pair in the duality interval is given
by~\cite{Vogt1995}
\begin{equation}
\sigma^{iQ}(Q\bar Q) = f^{iQ}_{Q \bar Q /p} \cdot P_{iQ} \cdot
  \sigma^{\it inel}_{pp} \cdot \frac{1}{9} \frac{\mu^2}{4\hat{m}_Q},
\end{equation}
where $\mu \approx 0.2$ GeV denotes the soft interaction scale parameter,
$f^{iQ}_{Q \bar Q /p}$ is the fragmentation ratio of the $Q \bar Q$ pair
written as
\begin{equation}
f^{iQ}_{Q \bar Q /p} = \int_{4m_Q^2}^{4m_H^2} dM_{Q \bar Q}^2
  \frac{dP_{iQ}}{dM_{Q \bar Q}^2} \,\, \bigg/
  \int_{4m_Q^2}^{s} dM_{Q \bar Q}^2 \frac{dP_{iQ}}{dM_{Q \bar Q}^2},
\end{equation}
and the inelastic proton--proton cross section $\sigma^{\it inel}_{pp}$ in the
region of $\sqrt{s}\geq 100$ GeV is obtained by the approximation~\cite{Block}
\begin{equation}\label{inel}
\sigma^{inel}_{pp} = 62.59 \, \hat{s}^{-0.5} + 24.09
  + 0.1604 \, \ln (\hat{s}) + 0.1433 \, \ln^2 (\hat{s}) \, \,\, \text{mb},
\end{equation}
where $\hat{s} = s/2m_p^2$. At the \hbox{\rm AFTER@LHC} energy
$\sqrt{s} = 115$ GeV, one obtains $\sigma^{inel}_{pp} = 28.4$ mb.

\subsection{Double-charmonium production from $|uud c\bar cc\bar c\rangle$}

The double-charmonium production cross section $\sigma(c\bar c+c\bar c)$ from
the Fock state $| uud c \bar c c \bar c \rangle$ can be written obviously as
\begin{equation}\label{dicc}
\sigma^{icc}(c \bar c + c \bar c) = (f^{icc}_{c \bar c/p})^2 \, P_{icc} \,
  \sigma^{inel}_{pp} \, \frac{1}{9} \frac{1}{9} \frac{\mu^2}{4\hat{m}_c},
\end{equation}
where the fragmentation ratio $f^{iQ_1Q_2}_{Q \bar Q/p}$ is obtained as
\begin{equation}\label{fcc}
f^{iQ_1Q_2}_{Q \bar Q/p} = \int_{4m_Q^2}^{4m_H^2} dM_{Q \bar Q}^2
  \frac{dP_{iQ_1Q_2}}{dM_{Q \bar Q}^2} \,\, \bigg/
  \int_{4m_Q^2}^{s} dM_{Q \bar Q}^2 \frac{dP_{iQ_1Q_2}}{dM_{Q \bar Q}^2} .
\end{equation}
In this case ($Q=c$, $H=D$) we use $m_c \approx 1.3$ GeV for the mass of
$c$ quark, $\hat{m}_c = 1.5$ GeV for the effective transverse $c$-quark mass,
and $m_D = 1.87$ GeV for the mass of the $D$ meson. For the integrated
probability distribution we take the value
$P_{icc} \simeq 0.002$~\cite{Vogt1995}.

Combining Eqs.~(\ref{dicc}) and~(\ref{fcc}), we may expect the
double-charmonium production cross section to be
\begin{equation*}
\sigma^{icc}(c \bar c + c\bar c) \approx 1.5 \times 10^2 \, \text{pb}.
\end{equation*}
Analyzing the values of the radial wave functions at the
origin~\cite{Eichten}, one finds
\begin{eqnarray*}
\sigma(J/\psi + J/\psi) : \sigma(J/\psi + \psi(2S)) :
  \sigma(\psi(2S) + \psi(2S)) \approx 1:0.65:0.43
\end{eqnarray*}
Taking into account Eq.~(\ref{rhoJpsi}) and the generalization of
Eq.~(\ref{XQQ}),
\begin{equation}
\sigma(X_1+X_2) = \rho_{X_1} \rho_{X_2}\cdot \sigma(Q\bar Q+Q\bar Q),
\end{equation}
one obtains
\begin{eqnarray}
\sigma^{icc}(J/\psi + J/\psi) \approx 6.0 \, \text{pb} \nonumber \\
\sigma^{icc}(J/\psi + \psi(2S)) \approx 3.9 \, \text{pb} \nonumber \\
\sigma^{icc}(\psi(2S) + \psi(2S)) \approx  2.6 \, \text{pb} 
\end{eqnarray}

\subsection{Associated charmonium--bottomonium production from
$|uudc\bar cb\bar b\rangle$}

Following Refs.~\cite{Vogt1995_1,Vogt1995_2}, the associated
charmonium--bottomonium production cross section is given by
\begin{equation}\label{ccbb}
\sigma^{icb}(c \bar c + b \bar b) = f^{icb}_{c \bar c/p} \,
  f^{icb}_{b \bar b/p} \, P_{icb} \, \sigma^{incl}_{pp} \frac{1}{9}
  \frac{1}{9} \frac{\mu^2}{4 \hat{m}^2_b}\Bigg( \frac{\hat{m_c}}{\hat{m_b}}
  \frac{\alpha_s (M_{b \bar b})}{\alpha_s (M_{c \bar c})} \Bigg)^4 \,.
\end{equation}
Applying Eq.~(\ref{fcc}) to this case ($Q=b$, $H=B$) we use
$m_b \approx 4.2$ GeV for the mass of the $b$ quark, $\hat{m}_b = 4.6$ GeV for
the effective transverse $b$-quark mass, and $m_B = 5.3$ GeV for the mass of
the $B$ meson. The value of $P_{icb}$ is unknown at this moment but we assume
it to be approximately equal to $P_{icc}$. Finally, we calculate the
associated charmonium--bottomonium production cross section to be
\begin{equation}
\sigma^{icb}(c \bar c + b \bar b) = 0.35 \, \text{pb}.
\end{equation}
In this section we calculate only the production cross section for the ground
states,
\begin{equation}
\sigma^{icb}(J/\psi + \Upsilon(1S)) \approx 14 \, \text{fb}.
\end{equation}

\subsection{Double-bottomonium production from $|uudb\bar bb\bar b\rangle$}

We already have all ingredients for the calculation of the production cross
section of the double-bottomonium states except for
$P_{ibb} = (\hat{m}_c / \hat{m}_b)^2 \cdot P_{icb}$, so the numerical value
will be
\begin{equation}
\sigma^{ibb}(b \bar b + b \bar b) = 0.03 \, \text{pb},
\end{equation}
and the cross sections for the particular double-botomonium states are given
by
\begin{eqnarray}
\sigma^{ibb}(\Upsilon(1S) + \Upsilon(1S)) \approx 1.2\, \text{fb} \nonumber \\
\sigma^{ibb}(\Upsilon(1S) + \Upsilon(2S)) \approx 0.6\, \text{fb} \nonumber \\
\sigma^{ibb}(\Upsilon(2S) + \Upsilon(2S)) \approx 0.3\, \text{fb}
\end{eqnarray}

\begin{figure}[ht]
\includegraphics[scale=0.66] {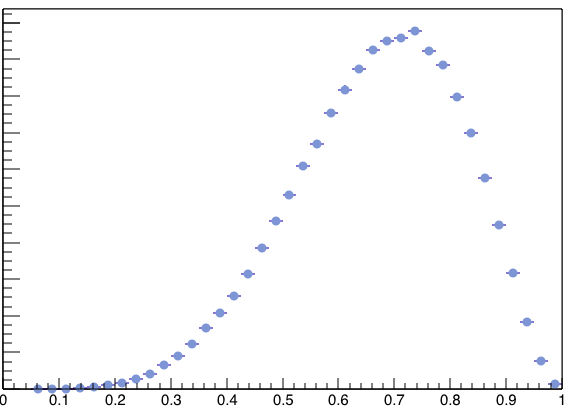}
\caption{\label{fig:shape}The histogram shows the $x_F$ distribution of the
  $J/\psi$ pair due to the double intrinsic heavy quark mechanism in
  arbitrary units.}
\end{figure} 
 
\section{Conclusions}
 In this paper we investigated the contribution of the double intrinsic heavy
quark mechanism to the production of a quarkonium pair. It is clear that
Single Parton Scattering (SPS) and Double Parton Scattering (DPS) provide the
main contributions to the double quarkonium production cross section. However,
both these contributions are vanishing fast with increasing Feynman parameter
$x_F$. On the other hand, the contribution from the double intrinsic heavy
quark mechanism mainly grows with $x_F$ (see Fig.~\ref{fig:shape}). If one
considers proton--proton collisions in the center-of-mass frame, one can
distinguish between charm production at positive $x_F$ coming from the
intrinsic heavy in the beam proton and negative $x_F$ coming from the
intrinsic heavy in the nucleons of the target. As it shown in
Ref.~\cite{Shao}, the DPS contribution starts at $x_F = -0.5$. This is the
region where the double intrinsic heavy quark mechanism is from the target and
contributes on the average. On the other hand, the double intrinsic charm
becomes the leading production mechanism at high $x_F$,
$\langle x_{\psi \psi} \rangle \simeq 0.64$~\cite{Vogt1995}.

\begin{figure}[ht]
\includegraphics[scale=1.0] {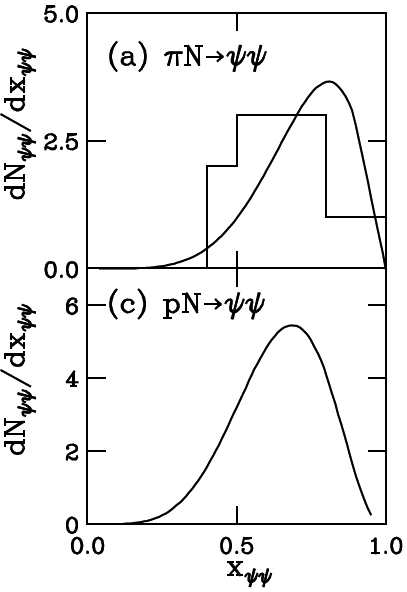}
\caption{\label{fig:na3_data}$x_F$ distribution for (a) $\pi N \to \psi\psi$
  and (c) $p N \to \psi\psi$ ($x_F\in[0,1]$). The plots are taken from
  Ref.~\cite{Vogt1995}. Shown are NA3 $\pi^{-} N$ data at $150$ and $280$
  GeV/c~\cite{NA31982} (histograms) and estimates of the intrinsic heavy
  quark mechanism (solid curve).} 
\end{figure}

Another interesting aspect to be discussed is
$\sigma(J/\psi+J/\psi)/\sigma(J/\psi)$. The only result for this ratio
with access to high values of $x_F$ was provided by the NA3 experiment and was
found to be $(3 \pm 1) \times 10^{-4}$ with $150$ and $280$ GeV/c
pion~\cite{NA31982} and $400$ GeV/c proton~\cite{NA31985} beams. The same
ratio measured by the LHCb Collaboration is found to be
$(5.1\pm 1.0\pm 0.6^{+1.2}_{-1.0})\times 10^{-4}$~\cite{LHCb_psi}.
This result can be interpreted wrongly as non-evidence for the intrinsic heavy
quark mechanism. However, the traditional $q\bar q$ annihilation mechanism
and the leading gluon-gluon fusion mechanism for LHCb are not in good
agreement with the NA3 data (cf.\ the discussion in Ref.~\cite{Vogt1995_2})
which shows that perturbative QCD can explain neither the NA3 cross section
nor the $x_F$ distribution. Compared to this, the double intrinsic heavy
quark mechanism reproduces $x_F$ dependencies very well~\cite{Vogt1995}, at
least for the case measured by NA3, namely the case of pion-nucleon
scattering~\cite{NA31982} (cf.\ Fig.~\ref{fig:na3_data}).

Current experimental knowledge does not give us much information about the
main contribution of the double intrinsic heavy quark mechanism. In our
calculations we use $P_{icc}$ from data with low statistics and provide other
formal assumptions. However, the key feature of \hbox{\rm AFTER@LHC} is the
access to high Feynman-$x$. Therefore, the measurement of the double
quarkonium production can provide more accurate data and shed more light on
the role of the intrinsic heavy quark mechanism.

\subsection*{Acknowledgments}
This work was supported by the Estonian Research Council under Grant
No.~IUT2-27.


\begin{thebibliography}{00}

\bibitem{ATLAS}
  G.~Aad {\it et al.} [ATLAS Collaboration],
  JHEP {\bf 1404}, 172 (2014)

\bibitem{ATLAS2}
  G.~Aad {\it et al.} [ATLAS Collaboration],\\
  Eur.\ Phys.\ J.\ C {\bf 75}, no. 5, 229 (2015)

\bibitem{LHCb}	
  R.~Aaij {\it et al.} [LHCb Collaboration],
  JHEP {\bf 1206}, 141 (2012);\\
  Addendum: [JHEP {\bf 1403}, 108 (2014)]

\bibitem{LHCb_psi}
  R.~Aaij {\it et al.} [LHCb Collaboration],
  Phys.\ Lett.\ B {\bf 707}, 52 (2012)

\bibitem{DZero}
  V.M.~Abazov {\it et al.} [D0 Collaboration],\\
  Phys.\ Rev.\ D {\bf 90}, no. 11, 111101 (2014)

\bibitem{CMS}
  V.~Khachatryan {\it et al.} [CMS Collaboration],
  JHEP {\bf 1409}, 094 (2014)

\bibitem{Brodsky2013}
  S.J.~Brodsky, F.~Fleuret, C.~Hadjidakis and J.P.~Lansberg,\\
  Phys.\ Rept.\  {\bf 522}, 239 (2013)
	
\bibitem{Lansberg2012}
  J.P.~Lansberg, S.J.~Brodsky, F.~Fleuret and C.~Hadjidakis,\\
  Few Body Syst.\ {\bf 53}, 11 (2012)

\bibitem{Lansberg2014}
  J.P.~Lansberg {\it et al.},
  EPJ Web Conf.\ {\bf 66}, 11023 (2014)

\bibitem{Rakotozafindrabe}
  A.~Rakotozafindrabe {\it et al.},
  PoS DIS {\bf 2013}, 250 (2013)

\bibitem{Brodsky2015}
  S.J.~Brodsky, A.~Kusina, F.~Lyonnet, I.~Schienbein, H.~Spiesberger
  and R.~Vogt,
  Adv.\ High Energy Phys.\ {\bf 2015}, 231547 (2015)

\bibitem{Shao}
  J.~P.~Lansberg and H.~S.~Shao,
  Nucl.\ Phys.\ B {\bf 900}, 273 (2015)

\bibitem{NA31982}
  J.~Badier {\it et al.} [NA3 Collaboration],
  Phys.\ Lett.\ {\bf 114B}, 457 (1982)

\bibitem{NA31985}
  J.~Badier {\it et al.} [NA3 Collaboration],
  Phys.\ Lett.\ {\bf 158B}, 85 (1985)

\bibitem{SELEX2002}
  M.~Mattson {\it et al.} [SELEX Collaboration],\\
  Phys.\ Rev.\ Lett.\ {\bf 89}, 112001 (2002)

\bibitem{Mattson}
  M.~Mattson, Ph.D.\ thesis, Carnegie Mellon University, 2002

\bibitem{SELEX2005}
  A.~Ocherashvili {\it et al.} [SELEX Collaboration],\\
  Phys.\ Lett.\ B {\bf 628}, 18 (2005)

\bibitem{Vogt1995}
  R.~Vogt and S.J.~Brodsky,
  Phys.\ Lett.\ B {\bf 349}, 569 (1995)

\bibitem{Koshkarev}
  S.~Koshkarev and V.~Anikeev,
  arXiv:1605.03070 [hep-ph]

\bibitem{Brodsky1984}
  S.~J.~Brodsky, J.C.~Collins, S.D.~Ellis, J.F.~Gunion and A.H.~Mueller,\\
  DOE/ER/40048-21 P4, SLAC-PUB-15471

\bibitem{Franz}
  M.~Franz, M.V.~Polyakov and K.~Goeke,
  Phys.\ Rev.\ D {\bf 62}, 074024 (2000)

\bibitem{Brodsky1980}
  S.J.~Brodsky, P.~Hoyer, C.~Peterson and N.~Sakai,\\
  Phys.\ Lett.\  {\bf 93B}, 451 (1980)

\bibitem{Brodsky1981}
  S.J.~Brodsky, C.~Peterson and N.~Sakai,
  Phys.\ Rev.\ D {\bf 23}, 2745 (1981)

\bibitem{Amundson}
  J.F.~Amundson, O.J.P.~Eboli, E.M.~Gregores and F.~Halzen,\\
  Phys.\ Lett.\ B {\bf 372}, 127 (1996)

\bibitem{Humpert}
  B.~Humpert, P.~M\'ery,
  Z.\ Phys.\ C {\bf 20}, 83 (1983)

\bibitem{Eichten}
  E.J.~Eichten and C.~Quigg,
  Phys.\ Rev.\ D {\bf 52}, 1726 (1995)

\bibitem{Block}
  M.M.~Block and F.~Halzen,
  Phys.\ Rev.\ D {\bf 86}, 014006 (2012)

\bibitem{Vogt1995_1}
  R.~Vogt and S.J.~Brodsky,
  Nucl.\ Phys.\ B {\bf 438}, 261 (1995)

\bibitem{Vogt1995_2}
  R.~Vogt,
  Nucl.\ Phys.\ B {\bf 446}, 159 (1995)

\end{thebibliography}
\end{document}